\documentclass[preprint,prd,showpacs,amsmath,amssymb]{revtex4}

\usepackage{graphicx}
\usepackage{dcolumn}
\usepackage{bm}

\begin{document}
\title{%
Sensitivity of cosmic-ray experiments to ultra-high-energy photons:
reconstruction of the spectrum and limits on the superheavy dark matter
}
\author{O.E.~Kalashev}
\author{G.I.~Rubtsov}
\author{S.V.~Troitsky}
\affiliation{Institute for Nuclear Research of the Russian Academy of
Sciences,\\
60th October Anniversary Prospect 7a, Moscow 117312 Russia
}
\begin{abstract}
We estimate the sensitivity of various experiments detecting
ultra-high-energy cosmic rays to primary photons with energies above
$10^{19}$~eV. We demonstrate that the energy of a primary photon may be
significantly (up to a factor of $\sim 10$) under- or overestimated for
particular primary energies and arrival directions.
We consider distortion of the reconstructed cosmic-ray spectrum for the
photonic component. As an example, we use these results to constrain the
parameter space of models of superheavy dark matter by means of both the
observed spectra and available limits on the photon content. We find that
a significant contribution of
ultra-high-energy particles (photons and protons) from decays of superheavy
dark matter is allowed by all these constraints.
\end{abstract}

\maketitle

\section{Introduction}
Recent studies~\cite{AGASA_Risse,A+Y,Ylim,Auger_sdlim} put strong limits
on the presence of photons among primary particles of
ultra-high-energy (UHE, $\gtrsim
10^{19}$~eV) cosmic rays (CR). However, while at energies $\sim
10^{19}$~eV current gamma-ray limits ($<2\%$ of the integral flux of
cosmic particles at 95\% CL~\cite{Auger_sdlim}) are very restrictive, the
best limit at $10^{20}$~eV allows (at 95\% CL) as much as 36\% of primary
gamma rays~\cite{A+Y}. At the same time, the reconstruction of the UHECR
spectrum often relies on a general assumption of hadronic primaries.
This assumption is explicit
in Monte-Carlo simulations (AGASA~\cite{AGASA_Eest}, Telescope
Array~\cite{TAsurface}) and implicit in methods which use calibration
relations obtained for the bulk of the lower-energy events which are mostly
hadronic (Yakutsk~\cite{PravdinICRC2005}, Pierre
Auger~\cite{Auger_sd_eest}). Well justified at $10^{19}$~eV, this approach
may lead to systematic distortions of the spectrum in the very interesting
energy range $\gtrsim 10^{20}$~eV, where a significant fraction of gamma
rays is allowed. Because of this systematics, models which predict primary
photons of these energies should be tested with great care.

The purpose of this paper is twofold. First,
we allow for a hypothetical contribution of primary photons which is
consistent with all experimental limits and study its possible effect on
the derivation of the spectrum. Second, we consider particular theoretical
models which predict such a contribution and constrain them with the
simultaneous account of the spectrum and of photon limits.  Though
ultra-high energy photons have not been observed by now, they are expected
to be seen among secondary products of the Greisen--Zatsepin--Kuzmin
\cite{g,zk} reaction (see e.g.\ Ref.~\cite{GZKphotons}). They are also
predicted in exotic hypothetical top-down models of UHECR origin (see
Refs.~\cite{siglreview,KachDMrev} for reviews), notably in the superheavy
dark-matter (SHDM) models.

In this paper, we give a quantitative analysis of reconstruction of
the spectrum by various experiments in the presence of a fraction of
gamma-ray primaries. This analysis is obligatory when exotic scenarios
of UHECR origin are constrained: theoretical predictions for the
photon fraction depend on the normalization of the exotic contribution
to the spectrum.

Air showers induced by primary photons differ
significantly from the bulk of hadron-induced events (see e.g.\
Ref.~\cite{RisseRev} for a recent review).
There are two competitive effects responsible for the diversity of
showers induced by primary photons.  First, due to the Landau,
Pomeranchuk \cite{LP} and Migdal \cite{M}~(LPM) effect the
electromagnetic cross-section is suppressed at energies
$E~>~10^{19}$~eV. The LPM effect leads to the delay of the first
interaction and the shower arrives to the ground level under-attenuated.
Another effect is the $e^\pm$ pair production due to photon interaction
with the geomagnetic field above the atmosphere. Secondary electrons
produce gamma rays by synchrotron radiation generating a cascade in
the geomagnetic field. The probability of this effect is proportional to
the square of the product of photon energy and perpendicular component of
geomagnetic field. The shower development therefore depends on both zenith
and azimuthal angles of photon arrival direction. If the effect is strong
enough, the particles enter the atmosphere with energies below the LPM
threshold. As a result, not only the shower development differs from
that of an average hadronic shower but also this difference is strongly
direction-dependent.

The energy reconstructed by an experiment may therefore differ
significantly from the true energy of the primary photon. In addition,
acceptance of fluorescence detectors for photons may differ from that for
hadronic primaries which is assumed in the spectral calculation. The
difference in the reconstructed energy and acceptance should be accounted
for individually for each experiment.
In this note, we estimate quantitatively this difference in
the energy reconstruction and discuss its possible effect on the
spectrum and implications for constraining the SHDM models.

\begin{figure}
\begin{center}
  \includegraphics[width=0.7 \columnwidth]{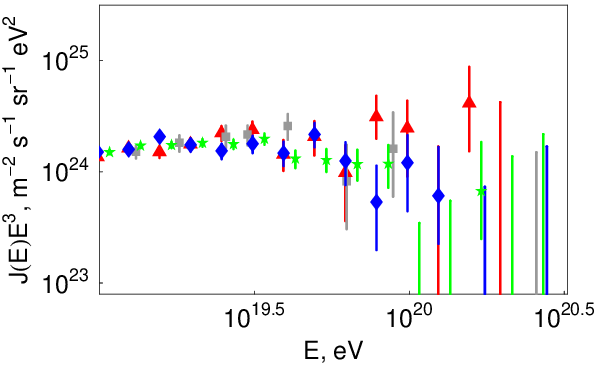}
\end{center}
\caption{
\label{fig:4spec}
Spectra of AGASA~\cite{AGASA_eest} (red triangles),
HiRes~\cite{HiRes_spec} (blue diamonds), Yakutsk~\cite{Yakutsk_spec} (grey
boxes) and PAO \cite{Auger_spec} (green stars) for energies scaled
according to the best fit with protons from extragalactic sources
described in Sec.~\ref{sec:shdm}.
}
\end{figure}

UHECR spectra measured by different
experiments are not in the mutual agreement. The disagreement is sometimes
attributed to systematic errors in energy
determination. Both the normalization of AGASA~\cite{AGASA_eest},
HiRes~\cite{HiRes_spec} and Yakutsk~\cite{Yakutsk_spec} spectra and the
position of the astrophysically motivated dip agree within this approach
\cite{Berezinsky_escale}. The spectrum observed by the Pierre
Auger Observatory (PAO) \cite{Auger_spec} agrees with the others in the
region above $10^{19}$~eV in the same assumption, see Fig.~\ref{fig:4spec}.
The energy rescaling is motivated by
discrepancies in different methods of energy estimation for hadronic
showers: for instance, the energy estimated by the surface
detector of PAO alone is about 30\% larger than the energy estimated in
the standard reconstruction procedure based on calibration to
fluorescence-detector data
\cite{Engel_cic}. For AGASA and Yakutsk the use of CORSIKA for
energy estimation leads to systematic shifts of energies downwards by
about 10-15\% and 40\%,
respectively~\cite{Teshima_ichep,Dedenko_yak_en,Dedenko_yak_confirm}. Note
that the rescaled spectra do not coincide at the highest energies ($E
\gtrsim 10^{20}$~eV); the discrepancy may be attributed either to
insufficient statistics or to the presence of energy-dependent
systematics, for instance of a non-standard component at the highest
energies. This question should be addressed in the future with more data
available.

\begin{figure}
\begin{center}
\includegraphics[width=0.8 \columnwidth]{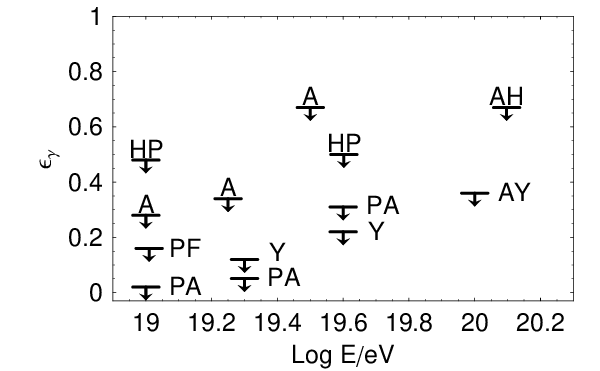}
\end{center}
\caption{
\label{fig:gamma-limits}
Upper limits (95\% C.L.) for the fraction $\epsilon_\gamma$ of gamma rays
in the integral flux of cosmic rays with energies higher than $E$:
Haverah Park~\cite{HP_lim} (HP), AGASA
\cite{AGASA_1stlim} (A), \cite{AGASA_Risse} (AH), AGASA and Yakutsk
\cite{A+Y} (AY), Yakutsk \cite{Ylim} (Y), Auger fluorescence
detector \cite{Auger_fdlim} (PF), Auger surface detector
\cite{Auger_sdlim} (PA). }
\end{figure}

Several limits on the fraction $\epsilon_\gamma$  of UHE photons in the
integral cosmic-ray flux have been set by several independent experiments.
They are summarized in Fig.~\ref{fig:gamma-limits}.
The most restrictive limits (95\% C.L.) are
$\epsilon _\gamma <0.36$ for
$E>10^{20}$~eV from the AGASA and Yakutsk
joint dataset~\cite{A+Y},
$\epsilon _\gamma <0.22$ for
$E>4\times 10^{19}$~eV from Yakutsk \cite{Ylim},
$\epsilon _\gamma <0.05$ for
$E>2\times 10^{19}$~eV
and
$\epsilon _\gamma <0.02$ for
$E> 10^{19}$~eV from the Auger surface detector
\cite{Auger_sdlim}. Even when the energy reconstruction of photons is
properly taken into account (which was done in the calculation of these
most restrictive limits), the limits on $\epsilon _\gamma$ may
depend on the uncertainty in the energy reconstruction of non-photonic
primaries, notably in the case of low statistics (see discussion and Fig.~2
in Ref.~\cite{Ylim}). A more stable quantity is the flux of gamma rays;
the Pierre Auger upper limit on the integral flux of photons above
$10^{19}$~eV is $3.8\times10^{-3}$~km$^{-2}$sr$^{-1}$yr$^{-1}$ at the 95\%
CL.
These photon limits may be used to constrain top-down models provided a
theoretical model for the top-down photon flux is given. In a
self-consistent analysis, the latter should be normalized to the observed
spectrum. This normalization requires in turn the account of the energy
reconstruction of photons which constitute a significant fraction of the
top-down flux. Below, we perform a joint analysis of the spectrum and
of photon limits for the SHDM models and constrain the space of two SHDM
parameters (mass and lifetime of the superheavy particle). Contrary to the
previous studies, most of which used either a naive AGASA normalization or
an overall (independent from the energy and arrival direction)
multiplicative correction for the reconstructed photon energies, our
results suggest that a significant fraction of cosmic rays from the SHDM
decays is allowed by all experimental constraints.

The rest of the paper is organized as follows. In Sec.~\ref{sec:sens} we
estimate sensitivity of four experiments (AGASA, HiRes, Pierre Auger and
Yakutsk) to the primary photon component. In Sec.~\ref{sec:shdm} we
consider an example of constraining SHDM parameters using primary spectra
and photon limits. Section~\ref{sec:concl} summarizes our results.

\section{Sensitivity to the photon component}
\label{sec:sens}
\begin{figure}
\begin{center}
\includegraphics[width=0.35 \columnwidth]{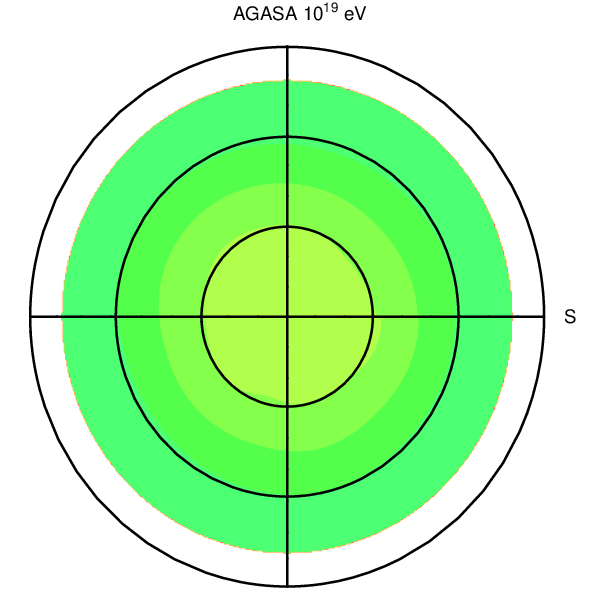}~%
\includegraphics[width=0.35 \columnwidth]{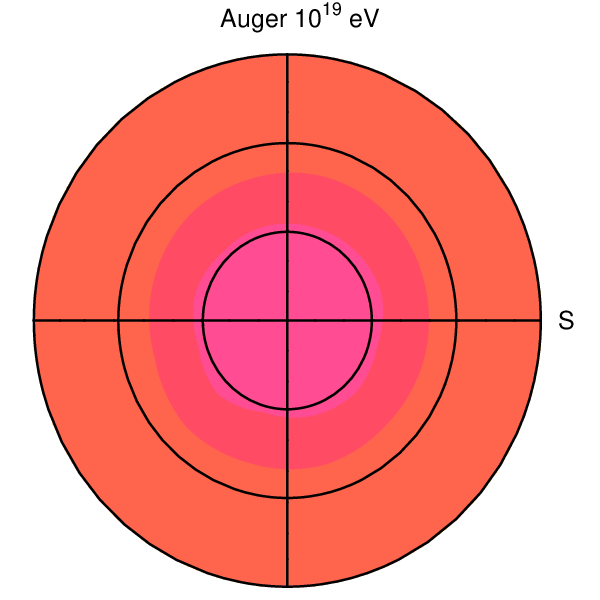}\\
\includegraphics[width=0.35 \columnwidth]{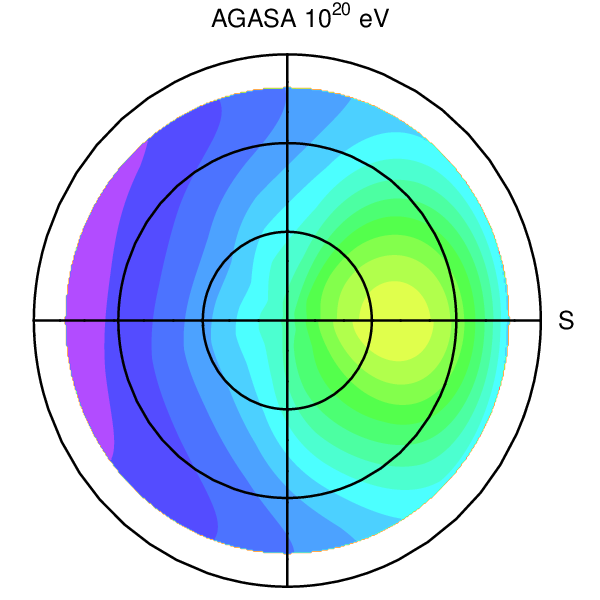}~%
\includegraphics[width=0.35 \columnwidth]{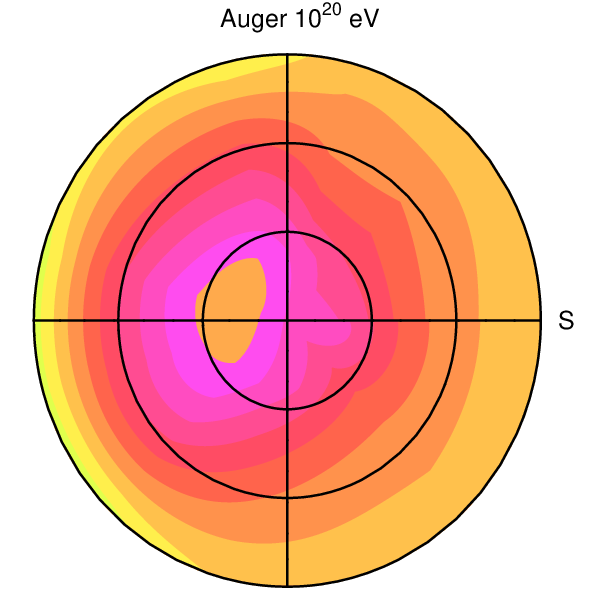}\\
\includegraphics[width=0.35 \columnwidth]{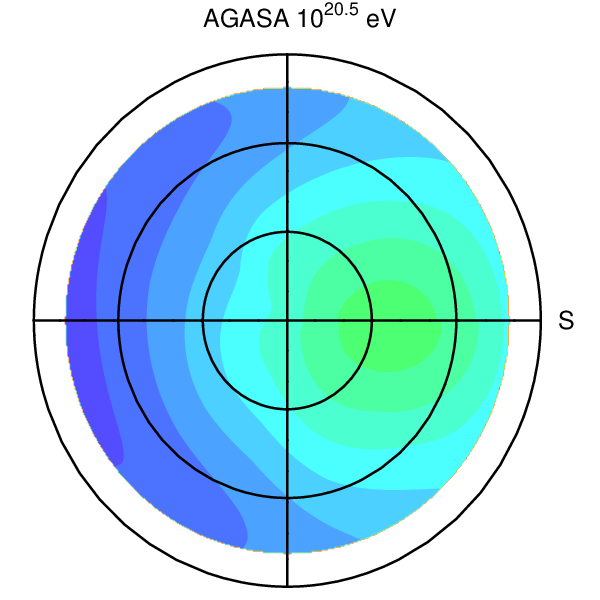}~%
\includegraphics[width=0.35 \columnwidth]{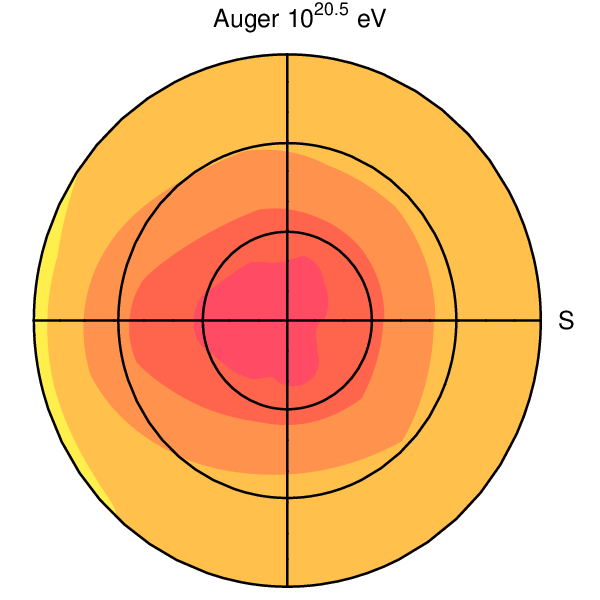}\\
\includegraphics[width=0.65 \columnwidth]{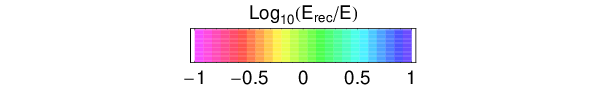}
\end{center}
\caption{
\label{fig:direc-PAO}
Energy overestimation factor for photon showers observed by AGASA (left
column) and by the surface detector of PAO as a function of energy (shown
for three values marked on the plots) and arrival direction (radial
coordinate: zenith angle, angle coordinate: azimuth; zenith is in the
center and South is to the right of the plots). The logarithmic colour code
is shown in the bottom panel. }
\end{figure}

To calculate the spectrum of photons reconstructed by a given experiment
it is important to account both for the energy estimation of a particular
photon and for the experiment's exposure to photons. We obtain
approximate estimates in the following way:

{\it AGASA.} The array has a geometrical exposure for hadronic primaries
with energies above $10^{18.5}$~eV. The probability to accept an event
by the ground detector depends only on the detector signal which is, for a
given arrival direction, in one-to-one correspondence with the
reconstructed energy of the event. Therefore, the exposure is geometrical
for photons with reconstructed energies above $10^{18.5}$~eV. To calculate
reconstructed energies for primary photons we run Monte-Carlo simulations
using CORSIKA 6.611~\cite{corsika} with GHEISHA~\cite{GHEISHA} as a
low-energy hadronic interaction model and EPOS 1.61~\cite{EPOS} as a
high-energy hadronic model. Since the hadronic component carries a small
fraction of energy of a photon-induced shower, dependence of the results
on the choice of hadronic model is negligible within our precision.
EGS4~\cite{Nelson:1985ec} is used to model electromagnetic
interactions and the PRESHOWER code~\cite{Homola:2003ru} is used to
account
for possible interactions of the primary photons with the geomagnetic
field. The reconstructed energy for primary photons is calculated by means
of the standard AGASA procedure~\cite{AGASA_Eest} using the detector
response obtained from the GEANT simulations in Ref.~\cite{Sakaki}.
The results of the analysis are presented in Fig.~\ref{fig:direc-PAO},
left column.
Photon-induced showers penetrate deeply and are therefore younger when
they reach the surface detector, as compared to the hadronic ones. This
fact results in overestimation of the primary energy because of a bias in
the attenuation correction (which was fit to the bulk of -- presumably
hadronic -- showers by means of the constant intensity cuts method). On
average, energies of showers with $E>10^{19}$~eV are overestimated by a
factor of $\sim 2$, but for particular energies and arrival directions
where the geomagnetic cascade does not compensate the
LPM suppression, the overestimation may reach a
factor of ten. We note that in Ref.~\cite{A+Y}, energies of individual
highest-energy AGASA events have been estimated within the primary-photon
assumption, so the corresponding limit on the gamma-ray fraction does not
suffer from this problem.

{\it HiRes.} The exposure of the HiRes fluorescent detector for the
primary photons was calculated in Ref.~\cite{HiRes_acc} and has been
found to be almost twice smaller than the exposure for protons.  The
reason is the reduced efficiency of reconstruction of deep showers, so
that a significant part of them does not pass strict quality cuts
determined for the spectrum-related studies (e.g. the maximum of the
shower development fully seen). Energies reconstructed by the
fluorescence method for different primary particles differ only by the
contribution of particles not taking part in the electromagnetic
cascade. The correction is calculated in Ref.~\cite{Pierog_ICRC}; its
application for the gamma-ray showers results in primary photons
energy overestimation by about 10\%, well within the systematic
uncertainties.

{\it Pierre Auger.} The surface detector of PAO also has the
geometrical exposure at the highest energies we are interested in. The
detector response of Auger water tanks is not publicly available and
therefore we use $S(1000)$ values for photon-induced showers without
geomagnetic preshower from Fig.~3 of
Ref.~\cite{Billoir_s1000gamma}. We separately perform preshower
simulations using the CORSIKA PRESHOWER module for El Nihuil location
and use data from Ref.~\cite{Billoir_s1000gamma} for secondary
photons. Finally we reconstruct the primary energy using formulae of
Ref.~\cite{Auger_sd_eest}.  The results of our analysis are presented
in Fig.~\ref{fig:direc-PAO}, right column. It turns out that the
photon primary energies are underestimated (for the spectrum
derivation) by the PAO surface detector by the factor of four in
average. The underestimation may reach an order of magnitude for
particular energies and arrival directions. The physical reason for
the photon energy underestimation is hypersensitivity of the water
tanks to the muon component of the shower, while the latter is
strongly suppressed in photon-induced showers. A completely different
energy reconstruction procedure, which assumes primary gamma rays, has
been applied~\cite{Auger_sdlim} for the calculation of the photon
limits.  The latters are therefore insensitive to this problem.

{\it Yakutsk.} The exposure of the Yakutsk EAS array is also geometrical.
The spectrum below $10^{19}$~eV is obtained using a small
subarray~\cite{Yakutsk_spec}, so the exposure depends on the energy in a
known way. To obtain reconstructed energies for the primary photons we use
the Monte-Carlo simulations (similar to those described above for AGASA)
and the Yakutsk detector response obtained from GEANT
simulations in Ref.~\cite{Dedenko:2004ng}. Qualitatively, the results are
very similar to those obtained for AGASA. The gamma limits of
Refs.~\cite{A+Y,Ylim} have been calculated with the account of the energy
reconstruction for primary photons.

\begin{figure}
\centerline{
\includegraphics[width=0.8 \columnwidth]{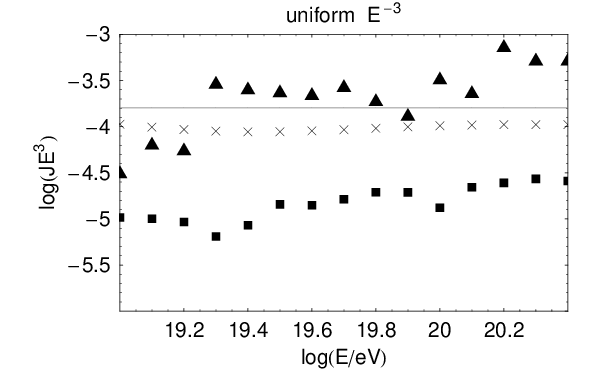}
}
\caption{
\label{fig:relative3}
The spectra reconstructed by different experiments (triangles: AGASA,
crosses: HiRes, boxes: Auger) for thrown isotropic photon flux $E^{-3}$
(gray line). }
\end{figure}

The results of the analysis for various experiments are illustrated in
Figs.~\ref{fig:direc-PAO}, \ref{fig:relative3}.

With the statistics presently available it is not
possible to explain the difference in spectra at the highest energies by
means of the photon component. Our consideration nevertheless illustrates
that the presence of a non-standard component might influence the
interpretation of experimental results.

\section{Constraints on the SHDM parameters}
\label{sec:shdm}
As an example of application of our results,
we study how the systematics in the determination of the spectrum in the
presence of primary photons may affect constraints on the SHDM obtained
from the limits on the primary gamma rays.

The SHDM models predict very hard spectrum with a large fraction of
photons and therefore both the spectral shape and gamma limits can be
used to constrain the models.  We perform a joint fit of spectra of
four experiments above $10^{19}$~eV with the sum of astrophysical and
SHDM components and obtain constraints on the parameters of SHDM
model.  The full spectral fit is performed as described below and 95\%
C.L.\ limits on the gamma-ray content listed in Sec.~\ref{sec:sens} are
imposed as theta-functional constraints. Since some of the gamma limits
are given in terms of the fraction, and not of the absolute flux, of
primary photons, they cannot be analysed independently of the spectra. For
the photon component, we use the spectrum reconstruction for each
experiment as described above, while for hadronic primaries, we consider
the energy scale as a parameter of fit individual for each experiment, see
details below. We take into account both photons and protons produced in
SHDM decays.

{\it The astrophysical contribution.}
We simulate propagation of cosmic rays from astrophysical sources using
the numerical code~\cite{propag} with a few modifications described
in Ref.~\cite{Gelmini:2007jy}.

The code uses the  kinematic-equation approach and
calculates the propagation of nucleons, stable leptons  and
 photons~\footnote{In the present study, we do not use the possibility of
 heavier nuclei primaries which is allowed by the code.} using the
 standard dominant processes (as is explained, for instance, in
 Ref.~\cite{siglreview}). For nucleons, it takes into account single and
 multiple pion production and $e^{\pm}$ pair production on the CMB,
 infrared/optical and radio backgrounds, neutron $\beta$-decays and the
 expansion of the Universe. For photons, the code includes  $e^{\pm}$ pair
 production, $\gamma + \gamma_B \rightarrow e^+ e^-$ and double $e^{\pm}$
 pair production $\gamma + \gamma_B \rightarrow e^+ e^-  e^+ e^- $,
 processes. For electrons and positrons, it takes into account inverse
 Compton scattering, $e^\pm + \gamma_B \rightarrow e^\pm \gamma$, triple
 pair production,  $e^\pm + \gamma_B \rightarrow e^\pm e^+ e^- $ ,  and
 synchrotron energy loss on extragalactic magnetic fields (EGMF). The
 propagation of  the electron-photon cascade and nucleons are calculated
 self-consistently, namely secondary particles arising in all reactions
are  propagated alongside the primaries. The  hadronic interactions of
 nucleons are derived from the well established SOPHIA event
 generator~\cite{Mucke:1999yb}.

 UHE particles lose their energy in interactions with the photon
background, which consists of CMB,  radio, infra-red and optical (IRO) components,
as well as EGMF.  Protons are sensitive essentially to the CMB only, while for
UHE photons and nuclei the radio and IRO components are respectively
important. Although the radio background is not yet
well known our conclusions do not depend strongly on the radio background assumed
since secondary photons from proton propagation are in any case
subdominant and their flux remains below the present limits (see
Ref.~\cite{Gelmini:2007jy} for more details; for SHDM the effect is also
not important because the SHDM flux is dominated by the Milky-Way
contribution). We use estimates by Clark {\it et al.}~\cite{clark} for
extragalactic radio background and model of Ref.~\cite{Stecker:2005qs} for
the IRO background component. The IRO background is only important to
transport the energy of secondary photons in the cascade process from the
0.1 - 100 TeV energy range  to the 0.1-100 GeV energy range observed by
EGRET,  and thus the resulting flux in the energy range of our interest is
not sensitive to details of the IRO background models.

We assume pure proton composition at injection and take the spectrum of an
individual UHECR source to be of the form:
\begin{equation}
F(E) = f E^{-\alpha} \Theta (E_{max} -E),  \label{proton_flux}
\end{equation}
where $f$ provides the flux normalization, $\alpha$ is the spectral  index and
$E_{\rm max}$  ($E_{\rm max}$) is the maximum energy to which protons
can be accelerated at the source.

We assume the standard cosmological model with the Hubble constant
$H=70$~km~s$^{-1}$~Mpc$^{-1}$, the dark energy density (in units of the
critical density) $\Omega_{\Lambda}= 0.7$ and a matter density
$\Omega_{\rm m}=0.3$. We define total source density in this model as
\begin{equation}
n(z) = n_0 (1+z)^{3+m_z}~
\Theta (z_{\max}-z) \Theta (z-z_{\min}) \,,\label{sources}
\end{equation}
where $m_z$  parameterizes the source density evolution, in such a  way
that $m_z=0$ corresponds to non-evolving sources with constant density
per comoving volume,  and $z_{\min}$ and $z_{\max}$ are respectively
the redshifts of the closest and most distant sources. In this paper we use
a fixed value of $z_{\max}=3$.

{\it The SHDM contribution.}
Decays of the SHDM particles may be described in a more or less
model-independent way because the most important physical phenomenon
of relevance is hadronization which involves light particles and is
well understood. Denote $x \equiv \frac{2E}{M_X}$, where $E$ is the
energy of a decay product of the SHDM particle with mass $M_X$. Then
for $10^{-4} \lesssim x \lesssim 0.1$, spectra of the decay products
calculated by various methods~\cite{ABK1,DMspectra} are in a good
agreement with each other; moreover, the shape of the spectral curve
$\frac{dN}{dE}(x)$ does depend only mildly on
$M_X$~\cite{ABK1} and we may consider the dependence negligible. For
this study, we use the spectra of decay products from
Ref.~\cite{ABK1}.\footnote{We thank M.~Kachelrie\ss\ for providing
numerical tables of the functions caculated there.}

The SHDM decay rate is determined by the concentration $n_X$ and lifetime
$\tau _X$ of the SHDM particles, $\dot{n}_X=n_X/\tau _X$. The flux of
secondary particles at the Earth is then determined by
$$
j={\cal N} \frac{1}{\tau _X}\,\frac{dN}{dE},
$$
where
\begin{equation}
{\cal N}=\int \!d^3r \, \frac{n_X({\bf r})}{4 \pi  r^2}
\label{*}
\end{equation}
is the geometrical factor (${\bf r}$ is the radius-vector from the Earth;
though in principle one should integrate over the Universe and account for
relativity effects, in most interesting cases the dominant contribution
comes from the Galactic halo~\cite{DubTin}). The mass $M_X$ is subject to
cosmological limits (see e.g.\ Ref.~\cite{newKolb} and references
therein); the lifetime $\tau _X$ is much less restricted.

\begin{figure}
\centerline{
\includegraphics[width=0.8 \columnwidth]{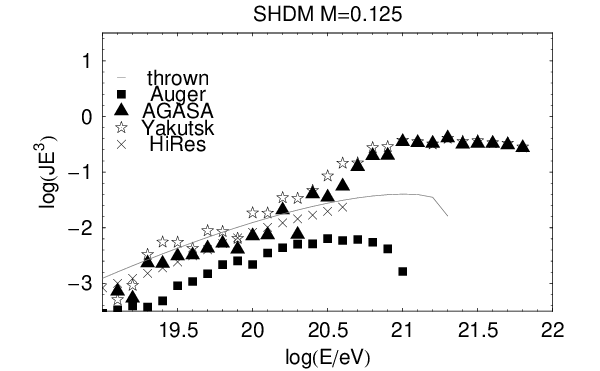}
}
\caption{
\label{fig:relative0.125}
Same as in Fig.~\ref{fig:relative3} but for the initial photon spectrum
predicted by decays of SHDM with $M_X=1.25 \times 10^{21}$~eV. }
\end{figure}

The flux is assumed to be a sum of two components,
one of which corresponds to the ``bottom-up'' contribution
while the second
one is due to the SHDM decays. While the former is assumed to be isotropic,
the latter is not because of non-central position of the Sun in the Milky
Way; we account for this anisotropy as described in Ref.~\cite{KKKST},
assuming the Navarro-Frank-White~\cite{NFW} dark-matter distribution and
with obvious account of the exposure (field of view) of particular
experiments. The account of the anisotropy reduces the difference in the
reconstructed spectra for the SHDM-related photons because the energy
underestimation by PAO is partially compensated by the larger flux of
photons seen in the Southern hemisphere, with an opposite effect for
AGASA, see Fig.~\ref{fig:relative0.125}.

{\it The fitting procedure.}
Up to the normalizations (depending on $\tau_X$ for the dark-matter part),
the spectra are determined by four parameters ($\alpha $, $E_{\rm max}$,
$m_z$ and $z_{\rm min}$) for the astrophysical part and by $M_X$ for the
SHDM part.
We scan over these parameters which are let to take their values on a
grid. For the astrophysical spectrum, we use the grid described in
Ref.~\cite{Gelmini:2007jy}; for $M_X$ we allow values $2^k
\times10^{22}$~eV for seven integer values of $k$, $-3\le k \le 3$. For
each point on the five-dimensional grid, we fit four experimental energy
spectra (AGASA~\cite{AGASA_eest}, Yakutsk~\cite{Yakutsk_spec},
HiRes~\cite{HiRes_spec} and PAO~\cite{Auger_spec}) with four
independent energy shifts representing energy-independent systematic
errors of the four experiments and with two overall normalization factors
(for the astrophysical and for the dark-matter parts), allowing these six
parameters to change continuously.

We fit binned numbers of events
detected by four experiments using the analog of $\chi^2$ for the Poisson
data described e.g.\ in Ref.~\cite{BakerCousins}. Technically, potential
systematic errors of the energy determination of hadronic primaries (fit
parameters) are taken into account as shifts of the theoretical curve and
not of the data. We fix the experiments' exposure and do not fix the total
number of the detected events.

Statistical errors in energy estimation are taken into account as
described in Ref.~\cite{Gelmini:2007jy}. They are assumed to be
Gaussian in logarithmic scale with widths 25\%, 20\%, 6\% and 17\% for
AGASA, HiRes, PAO and Yakutsk
respectively~\cite{PravdinICRC2005,AGASA_eest,HiRes_spec,Auger_spec}.

The goodness of fit is determined by the Monte-Carlo simulations as
described in Ref.~\cite{NR}.
We consider a fit as acceptable if its goodness
exceeds 0.05.

\begin{figure}
\begin{center}
\includegraphics[angle=0,width=0.8 \columnwidth]{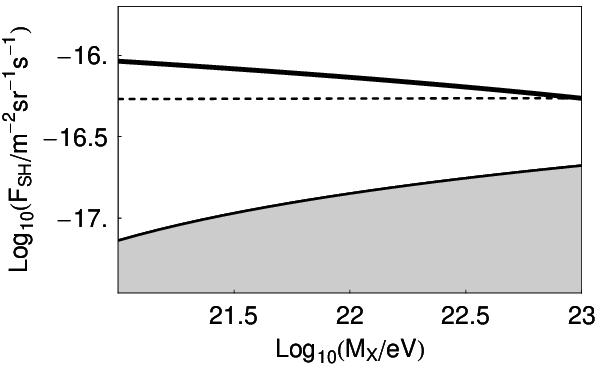}
\end{center}
\caption{
\label{fig:SHDMres}
Parameter space of the SHDM models: the total integral flux $F_{\rm
SH}$ of the cosmic rays from SHDM decays at energies $E>10^{20}$~eV
(inversely proportional to the lifetime $\tau_X$ of the SHDM particle
in particular models) versus the mass $M_X$ of the SHDM particle. The
area above the thick line is excluded by the spectral fits; the area
above the dashed line is excluded by the limit on $\epsilon _\gamma $
at $E>10^{20}$~eV \cite{A+Y}; the area above the thin line is excluded
by the limit on the gamma-ray flux at $E>10^{19}$~eV
~\cite{Auger_sdlim}.  The shadowed area is allowed by any
constraints.
}
\end{figure}

\begin{figure}
\begin{center}
\includegraphics[angle=0,width=0.48 \columnwidth]{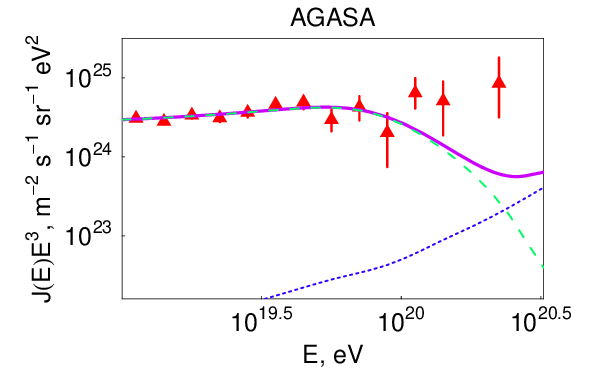}~%
\includegraphics[angle=0,width=0.48 \columnwidth]{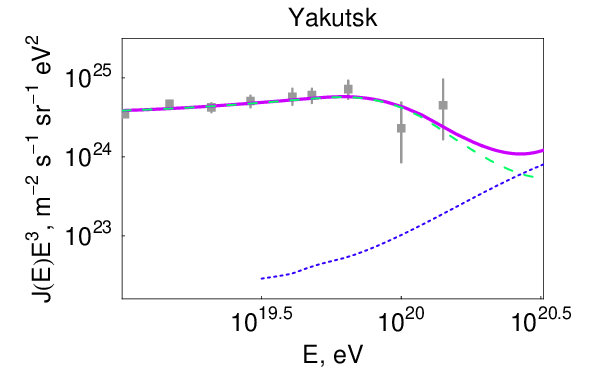}\\
\includegraphics[angle=0,width=0.48 \columnwidth]{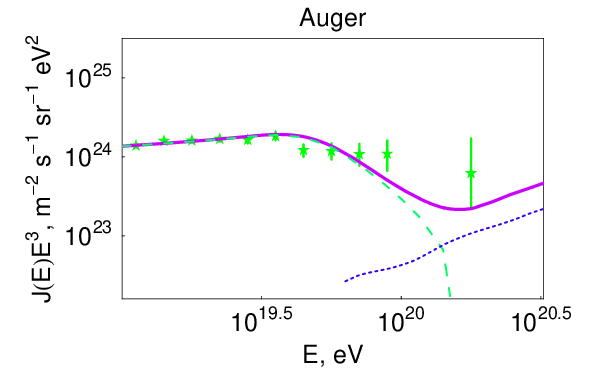}~%
\includegraphics[angle=0,width=0.48 \columnwidth]{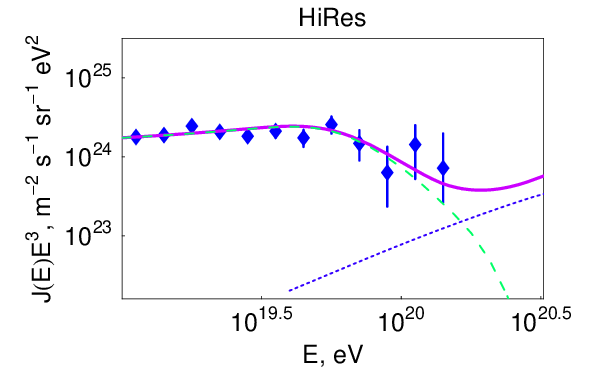}
\end{center}
\caption{
\label{fig:best-fit-spec}
The fit of the spectra observed by different experiments
with the ``extragalactic plus SHDM'' model for which the SHDM
contribution is maximal but all photon limits are satisfied. Symbols with
error bars represent experimental data points, thick lines represent the
fit, dashed lines -- the extragalactic component of the fit
function, dotted lines -- contribution of photons from the SHDM decays.
One and the same physical spectrum looks different for different
experiments because the energy reconstruction is taken into account: for
the hadronic component it is encoded in the energy shifts -- parameters of
 the fit; for the gamma-ray component it is obtained explicitly in
 Sec.~\ref{sec:sens}. }
\end{figure}

{\it Results.}
Resulting constraints on the SHDM parameters from the spectral fits and
photon limits are presented in
Fig.~\ref{fig:SHDMres}.
The spectral fits are equally good for models with and without SHDM
\footnote{%
The best fit without SHDM has a goodness of 0.19 and corresponds
to energy scaling factors of
0.92, 1.04, 0.70 and 0.60
for HiRes, PAO, AGASA and Yakutsk
respectively (astrophysical model $z_{\rm min}=0$,
$m_z=4$, $\alpha=2.45$, $E_{\rm max} = 1.28\times 10^{21}$~eV).
The best fit with SHDM (which satisfies all photon limits) has a goodness of
0.25 and corresponds to an SHDM model with $M_X = 2.5 \times 10^{21}$~eV
and energy shifts 0.95, 1.07, 0.72 and 0.61
for HiRes, PAO, AGASA and Yakutsk
respectively (astrophysical model $z_{\rm min}=0$,
$m_z=4$, $\alpha=2.45$, $E_{\rm max} = 6.4\times 10^{20}$~eV).
At (true) energies higher than $10^{20}$~eV, the SHDM-related component
(photons and hadrons) comprises 43\% of the total cosmic-ray flux for
the best fit parameters.}
and a significant part of the SHDM parameter space does not contradict the
observed spectra. As one might expect, the photon limits are more
restrictive, but our analysis demonstrates that, contrary to the common
lore, even they \textbf{ do not exclude SHDM} for a wide range of
parameters: more than one half of the cosmic-ray particles with $E \gtrsim
10^{20}$~eV may have their origin from the SHDM decays without violating
any of the experimental constraints.
In Fig.~\ref{fig:best-fit-spec},
we present gamma-ray and hadron spectra for one of the models with
maximal allowed SHDM contribution (similar spectra are obtained for
several models with slightly different parameters of the astrophysical
contribution).

\section{Conclusion}
\label{sec:concl}
Modern experiments have different sensitivities to the photon component
and this should be taken into account when testing particular models.
The AGASA experiment overestimated the energy of primary photons with
energies $E \gtrsim 10^{19}$~eV by a factor of 2 in average, though the
overestimation reaches a factor of $\sim 10$ for particular energies and
arrival directions. Contrary, the surface detector of the Pierre Auger
Observatory underestimates the energy of primary photons in this energy
range by a factor of $\sim 4$ in average, while underestimation by a
factor of $\sim 10$ happens for particular energies and arrival
directions. The HiRes detector overestimated the photon energies by a
factor of $\sim 1.1$, uniformly over arrival directions and well within
the systematic uncertainties. However, it had a significantly lower
exposure for primary photons than for primary hadrons.

One of the scenarios predicting a significant amount of primary UHE photons
is the superheavy-dark-matter model. We analyzed constraints on its
parameters from the observed spectra and limits on the photon content.
While the most restrictive photon limits~\cite{A+Y,Ylim,Auger_sdlim}
account for peculiarities in the energy reconstruction for photons, a
dedicated study was required and performed for
the spectral fits. A significant (more than one half of the flux at
$E>10^{20}$~eV) SHDM component is still allowed by all limits. Though
there seems no present need for the SHDM to explain the UHECR spectrum,
a large contribution of SHDM at the highest energies is not excluded and
may be further constrained (or validated) by future experiments. Among the
tests are measurements of the spectrum, studies of anisotropy and searches
for primary photons and neutrinos. Our study indicates that a significant
SHDM contribution is allowed for masses of dark-matter particles $M_X
\gtrsim 4\times 10^{22}$~eV, so one needs experiments with large aperture
at $E>10^{20}$~eV (e.g.\ JEM-EUSO~\cite{JEM-EUSO}) to test the shape of
the spectrum. At lower energies ($10^{19}$~eV to $10^{20}$~eV) the model
may be constrained by improving the gamma-ray limits and by precise
studies of the Galactic anisotropy, e.g.\ with the help of fluorescent
detectors (which treat uniformly both photon and hadron primaries).

As a final remark, we note that the example of photons should warn one
against naive tests of models predicting ``exotic'' primaries with the
experimental data. For instance, the correlations with BL Lac type objects
observed by HiRes \cite{BL:HiRes,HiRes:BL} require neutral primary
particles. If the latters were photons, apparent absence of correlations
in the preliminary data of the PAO surface detector \cite{PAO:BL} is
easily explained \cite{index} by underestimation of their energies as
compared to the bulk of hadronic primaries. With more exotic primary
particles, the analysis becomes even less trivial.

We are indebted to V.~Berezinsky, D.~Gorbunov, J.-M.~Fr\'ere, M.~Libanov,
M.~Pravdin, V.~Rubakov, D.~Semikoz, P.~Tinyakov and I.~Tkachev for
valuable discussions. This work was supported in part by the Russian
Foundation of Basic Research grants 07-02-00820 and 09-07-00388, by
Federal Agency for Science and Innovation under state contracts
02.740.11.0244 and 02.740.11.5092, by the grants of the President of the
Russian Federation NS-1616.2008.2, MK-1966.2008.2 (OK), MK-61.2008.2 (GR),
by the Dynasty Foundation (GR), by the Russian Science Support Foundation
(GR), by IISN and by the Belgian Science Policy (IAP VI-11).  Numerical
part of the work was performed at the computer cluster of the Theoretical
Division of INR RAS. GR and ST thank Service of Theoretical Physics, ULB,
for kind hospitality.


\end{document}